\documentclass[%
 reprint,
 amsmath,amssymb,
 aps,
]{revtex4-1}

\usepackage{graphicx}
\usepackage{dcolumn}
\usepackage{bm}

\begin{document}

\preprint{APS/123-QED}

\title{Cylindrical model for the dark matter halo of disk galaxies}

\author{Brian A. Slovick}
 \altaffiliation{brian.slovick@sri.com}
\affiliation{%
 Applied Optics Laboratory, SRI International, Menlo Park, CA USA 94025}%

\date{\today}

\begin{abstract}
A cylindrical model for the dark matter halo of disk galaxies is developed. At the center of the cylinder, in the plane perpendicular to the long axis, the rotation curve is constant for distances much less than the cylinder length and Keplerian at much greater distances. The rotation curve is equivalent to the spherical truncated flat (TF) profile, a model derived empirically from the radial velocity dispersion of the Milky Way dark halo. It is shown that an isothermal, self-gravitating cylinder of length 89 kpc can account for the observed radial velocity dispersion of the Milky Way dark halo with less mass than the NFW profile. Moreover, a cylindrical model of the Milky Way dark halo is consistent with free-streaming neutrinos of mass 1.1 eV.
\begin{description}
\item[PACS numbers]
95.35.+d, 98.80.-k, 98.52.Nr.
\end{description}
\end{abstract}


\maketitle


\section{\label{sec:level1}Introduction\protect\\}
To explain the flat rotation curves of spiral galaxies, the standard theory of galaxy formation invokes spherical halos of hypothetical cold dark matter (CDM) \cite{Navarro1996}. While the CDM model is in agreement with a number of astrophysical phenomena on large scales, on the scale of galaxies and below, several discrepancies arise between the CDM model and observation. Perhaps the most well known are the overabundance of small satellite galaxies in cosmological simulations \cite{Kauffmann1993} and the angular momentum problem associated with sharply-peaked dark matter distributions \cite{Moore1999}. The latter is exemplified in rotation curve measurements of low-luminosity, dark-matter dominated galaxies \cite{McGaugh2003}. In these systems, the linearity of the observed rotation curves implies a density profile that is nearly constant in the innermost regions \cite{Simon2005}, whereas numerical simulations with CDM invariably form cuspy dark matter distributions, having density profiles that vary as $\rho\propto r^{-1}$ in the central regions \cite{deBlok2001}. Cuspy dark matter halos are also in conflict with the mass distribution of the Milky Way inside the solar circle \cite{Binney2001}, evidence for maximum disks in spiral galaxies \cite{Binney2008}, and the rotational speeds of galactic bars \cite{Debattista2000}.

The shape of the dark matter halo plays a central role in determining the density distribution, total mass, and rotation curve of galaxies. Halo shapes are also important for understanding tidal streamers \cite{Law2009}, x-ray halos \cite{Pointecouteau2005}, and weak gravitational lensing data \cite{Hoekstra2004}. The shape of a dark matter halo is related to its gravitational field $\mathbf{g}$ through Gauss's law,
$$
\int \mathbf{g}\cdot d\mathbf{A}=-4\pi G M_{enc},
$$
where $M_{enc}$ is the mass enclosed within the surface area of the halo $A$, and $G$ is Newton's constant. Assuming that the gravitational field is constant over the surface, $\mathbf{g}$ can be removed from the surface integral to obtain
$$
g(r)=\frac{4\pi G M_{enc}}{A}.
$$
 The velocity of a circular orbit is then given by
\begin{equation} \label{eq1}
V_c(r)=[g(r)r]^{1/2}=\left[ \frac{4\pi G M_{enc}r}{A} \right]^{1/2}.
\end{equation}
Thus, the rotation curve is constant if $A$ varies linearly with radius, i.e., if the gravitational source is a cylinder. In contrast with a spherical halo, the cylindrical mass generates a flat rotation curve assuming all of the mass is contained within the Gaussian surface. This means that the rotation curve of a cylindrical dark matter halo may be flat in regions devoid of dark matter.

Cylindrical filaments arise naturally in simulations involving warm or hot dark matter \cite{Gao2007} because the large thermal velocities suppress density fluctuations below the free-streaming scale. Until recently, observations of the cosmological microwave background showed no evidence of the suppression of density fluctuations on small scales. But recent revelations regarding the sensitivity of the WMAP power spectra to the form of the WMAP beam and a subsequent analysis using microwave point sources for calibration \cite{Sawangwit2010} indicate that density fluctuations in the WMAP spectrum may not be as large as once believed, thus opening the door to alternative explanations of the classic rotation curve problem.

\section{\label{sec:level1}Thin Filament}

Consider a thin, one-dimensional dark matter filament of mass $M$ and length $l$. In the limit of weak gravitational fields, the gravitational potential $\Phi$ at a distance $r$ from the center of the filament and in a direction perpendicular to the long axis is given by \cite{Kellogg1954}
$$
\Phi(r)=-\frac{2GM}{l} \ln\left[{\frac{(l^2+4r^2)^{1/2}+l}{2r}}\right].
$$
A more general form of this potential has been derived previously to study the periodic orbits of bodies for which elongation is the gravitationally-dominant feature \cite{Riaguas1999}.

The objective is to derive the rotation curve of a thin disk located at the center of the filament, in the equatorial plane. From symmetry considerations, all gravitational forces parallel to the filament cancel, and the gravitational field in the plane of the disk is entirely in the radial direction with a magnitude
\begin{equation} \label{eq2}
g(r)=-\frac{d\Phi(r)}{dr}=-\frac{2GM}{r(l^2+4r^2)^{1/2}}.
\end{equation}
For distances much less than the filament length, the gravitational field obeys an inverse law. In this region, the filament can be regarded as an infinite cylinder, and the gravitational field is proportional to the ratio of the enclosed mass to the surface area of the cylindrical Gaussian surface, as in Eq. \ref{eq1}. At distances much greater than the filament length, the cylinder resembles a point mass and the gravitational field returns to the familiar inverse-square law.

The rotation curve of the thin dark matter filament will varies as
\begin{equation} \label{eq3}
V_c(r)=\frac{V_0}{(1+4r^2/l^2)^{1/4}}
\end{equation}
where
\begin{equation} \label{eq4}
V_0=\left( \frac{2GM}{l}\right)^{1/2}.
\end{equation}
The rotation curve is shown in Fig. 1. As expected, for distances much less than the filament length ($r<<l/2$), the circular velocity is constant with an amplitude $V_0$. At much greater distances, the filament resembles a point mass and the circular velocity returns to Keplerian form.

\begin{figure}
\includegraphics[width=88mm]{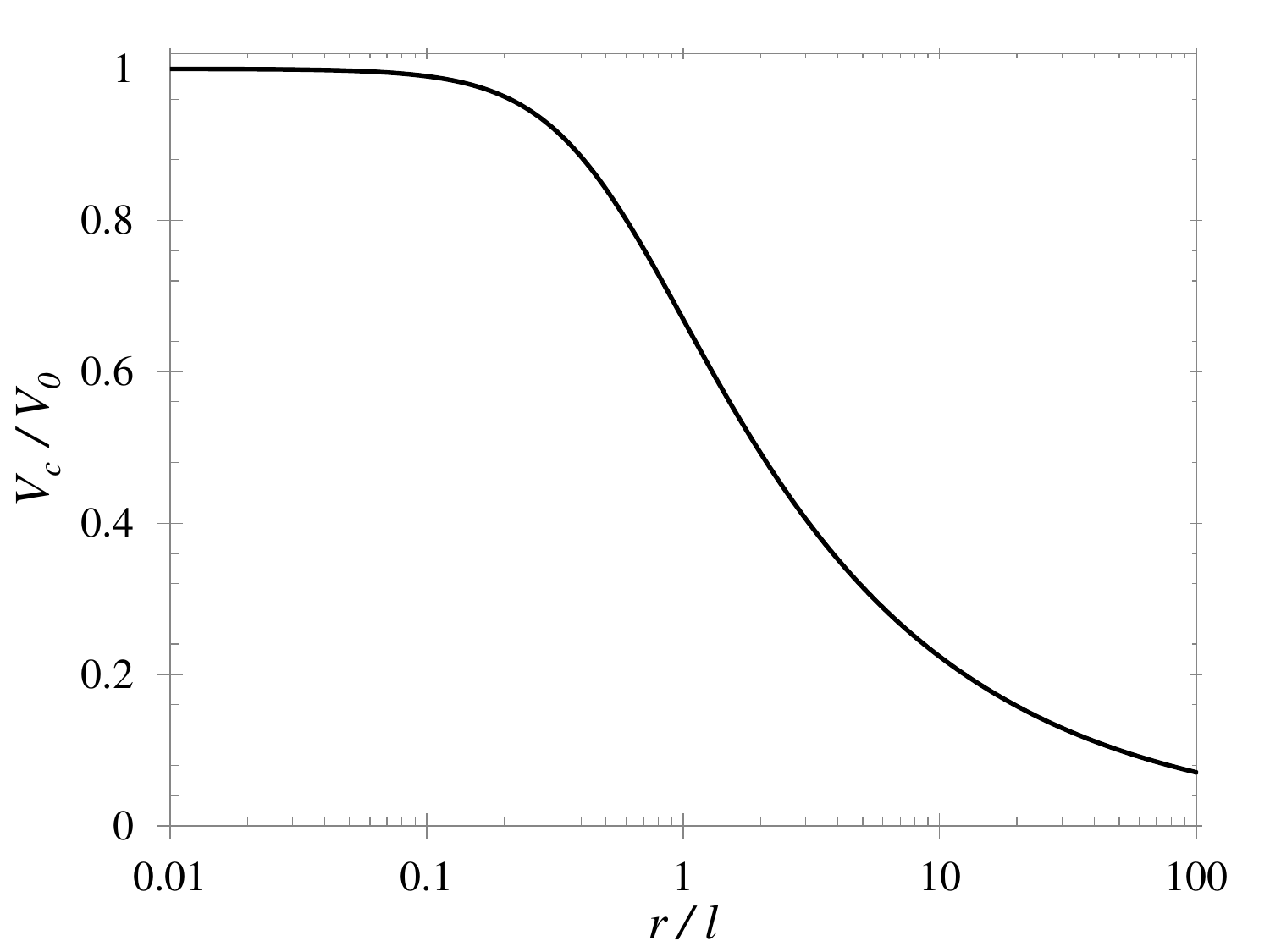}
\caption{\label{fig:epsart} Rotation curve for a one-dimensional cylinder of length $l$, from Eq. \ref{eq2}.}
\end{figure}

In the plane of the disk, the filament rotation curve in Eq. \ref{eq3} is equivalent to the truncated flat (TF) model \cite{Battaglia2005,Battaglia2006}. The TF model was introduced as an alternative to the well known Jaffe profile, and is distinguished by its abrupt decline in density at large distances. Originally, the TF model was derived empirically to determine the total mass of the Milky Way dark halo \cite{Wilkinson1999}. Under the assumption of constant velocity anisotropy, the TF model is known to provide a more accurate representation of the radial velocity dispersion of the Milky Way dark halo than both the pseudo-isothermal sphere and Navarro-Frenk-White (NFW) profile \cite{Battaglia2006}. The improvement is most evident at large distances from the center of the halo, where the TF model uniquely predicts a steep Keplerian decline. Earlier works employed the TF model to analyze the orbits of the Magellanic clouds \cite{Lin1982} and the kinematics of halo streams \cite{Lynden-Bell1995}.

\section{\label{sec:level1}Isothermal Filament\protect\\}
If the dark matter is warm or hot, which is necessary to sustain free-streaming filaments \cite{Gao2007}, the thermal velocities are not negligible. An exact analytical solution to the Jeans equation for an axisymmetric, isothermal, steady-state filament exists and has been used previously to determine the mass of large-scale filaments in galaxy red shift surveys \cite{Eisenstein1997}. For a filament that is translationally invariant along its symmetry axis, the mass density $\rho$ varies as
\begin{equation} \label{eq5}
\rho(r)=\frac{(2-\beta)^2\sigma^2}{2\pi GR^{\beta-4}}\frac{r^{-\beta}}{(R^{2-\beta}+r^{2-\beta})^2}
\end{equation}
where $\sigma$ is the velocity dispersion, $R$ is a scale parameter, and $\beta$ is the velocity anisotropy parameter. For an isothermal mass distribution, the velocity dispersion is related to the ensemble average of the radial velocities of the constituitive particles. Similar to the analysis of spherical systems, $\beta=1$ for purely radial orbits, $\beta=0$ corresponds to isotropic orbits, and $\beta \rightarrow \infty$ indicates tangential orbits.

By integration of Eq. \ref{eq5}, the filament mass enclosed within a radius $r$ is given by
\begin{equation} \label{eq6}
m(r)=2\pi l \int_0^r{dr' r'\rho(r')}=\frac{Mr^{2-\beta}}{R^{2-\beta}+r^{2-\beta}}.
\end{equation}
In contrast with the isothermal sphere and NFW profile, the total mass of the filament $M$ is finite, with a value
\begin{equation} \label{eq7}
M=\lim_{r\rightarrow \infty}m(r)=(2-\beta)\frac{\sigma^2 l}{G}.
\end{equation}
Therefore, the parameter $R$ may be interpreted as the radial distance where half of the total filament mass is enclosed. It follows from Eqs. \ref{eq7} and \ref{eq4} that the velocity dispersion is related to the amplitude of the rotation curve by
\begin{equation} \label{eq8}
V_0^2=2(2-\beta)\sigma^2.
\end{equation}

For distances much less than the filament length, the filament may be regarded as an infinite cylinder. According to Gauss's law, the gravitational field inside an infinite hollow cylinder vanishes because $M_{enc}=0$. Therefore, for distances much less than the filament length, the gravitational field depends only on the mass interior to $r$. In this approximation, the gravitational field of the isothermal filament is obtained by inserting $m(r)$ from Eq. \ref{eq6} into Eq. \ref{eq2}, and taking $r<<l/2$,
$$
g(r)=\frac{2G}{lr}m(r)=\frac{2G}{lr}\frac{Mr^{2-\beta}}{R^{2-\beta}+r^{2-\beta}}.
$$
Thus, the rotation curve of the isothermal, self-gravitating, steady-state filament is given by
\begin{equation} \label{eq9}
V_c(r)=V_0\left( \frac{r^{2-\beta}}{R^{2-\beta}+r^{2-\beta}} \right)^{1/2}.
\end{equation}

\begin{figure}
\includegraphics[width=88mm]{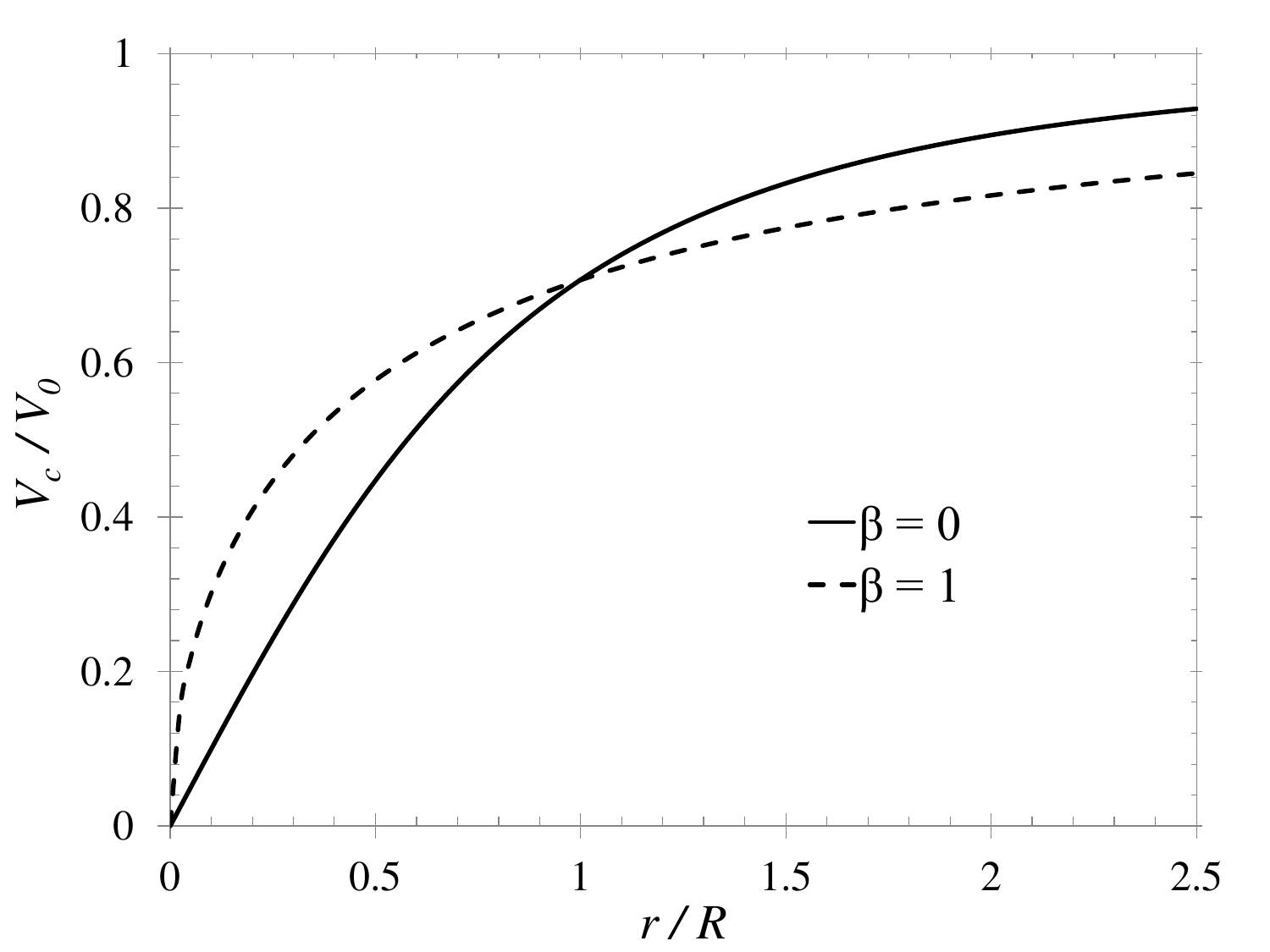}
\caption{\label{fig:epsart} Rotation curve of an isothermal, self-gravitating dark matter filament of radial scale length $R$ and velocity dispersion parameter $\beta$.}
\end{figure}

Figure 2 shows $V_c(r)$ for two values of $\beta$. Both rotation curves approach a constant value at large radius. From the assumptions leading to Eq. \ref{eq9}, the rotation curve remains flat in regions where $R<< r<<l/2$. For a filament with $R<<l/2$, the rotation curve valid for all distances can be obtained by inserting $m(r)$ into Eq. \ref{eq3}. The functional form of the inner rotation curve is determined by the velocity anisotropy parameter through the expression $V_c \propto r^{1-\beta/2}$. For a purely radial velocity ellipsoid ($\beta=1$), the mass density varies as $\rho \propto r^{-1}$ and the inner rotation curve is proportional to the square root of the radius, i.e., $V_c \propto r^{1/2}$. This behavior is characteristic of cuspy dark matter halos, e.g., the isothermal sphere and NFW profile \cite{Navarro1996}. If the velocity anisotropy ellipsoid is isotropic ($\beta=0$), the filament contains a constant density core and the rotation curve varies linearly with radius. Therefore, for an isothermal filament, a cuspy halo is avoided by reducing $\beta$, which according to Eq. \ref{eq8}, corresponds to reducing the velocity dispersion.

\section{\label{sec:level1}Milky Way Model}

To determine the dimensions of a filament that are consistent with the rotation curve of the Milky Way dark halo, consider the results of Battaglia et al. \cite{Battaglia2006}. In contrast with previous applications of the TF model \cite{Wilkinson1999,Sakamoto2003}, the approach of Battaglia et al. \cite{Battaglia2005,Battaglia2006} does not assume a spherical density distribution, and consequently, the results are applicable to the thin filament rotation curve in Eq. \ref{eq3}. In the following analysis, several empirical parameters are used to determine the properties of an isothermal filament. The empirial parameters and calculated filament properties are shown in Tables I and II.

\begin{table}
\caption{\label{tab:table2} Empircal parameters for the Milky Way \cite{Battaglia2006,Binney2008}.}
\begin{ruledtabular}
\begin{tabular}{cccccccc}
 $M_{halo}$&$\beta$&$M_{disk}$&$R_{disk}$&$r_{\odot}$&$V_c(r_{\odot})$\\
$[M_{\odot}]$& & $[M_{\odot}]$& [kpc]& [kpc]&[km s$^{-1}$]\\
\hline
\rule{0pt}{4ex}
5$\times10^{11}$&0.6&4.5$\times10^{10}$&2.5&8&220 \end{tabular}
\end{ruledtabular}
\end{table}

\begin{table}
\caption{\label{tab:table2}Parameters for the cylindrical model of the Milky Way Galaxy, calculated using the parameters in Table I.}
\begin{ruledtabular}
\begin{tabular}{cccccccc}
 $l$&$R$&$2R/l$&$\sigma$ &$\rho(r_{\odot})$\\
\hline
\rule{0pt}{4ex}
89 kpc& 8.6 kpc& 0.19 & 131 km s$^{-1}$& 0.19 GeV cm$^{-3}$\\
\end{tabular}
\end{ruledtabular}
\end{table}

Assuming constant velocity anisotropy, Battaglia et al. \cite{Battaglia2006} obtain a least squares fit to the observed radial velocity dispersion of the Milky Way dark halo for a TF model with $\beta=0.6$, consistent with the value observed in the solar neighborhood \cite{Sommer-Larsen1997}, and a relatively small halo mass of $5\times10^{11} M_{\odot}$, where $M_{\odot}$ is the solar mass. For the NFW profile, the same authors find a best fitting virial mass of $0.94-1.5\times10^{12} M_{\odot}$, significantly larger than the mass obtained with the TF model. In the context of the present work, this is readily explained by interpreting the TF model as describing a cylinder with finite mass. Setting the amplitude of the rotation curve to $V_0=220$ $\text{km s}^{-1}$, from Eq. \ref{eq4}, the corresponding filament length is $l=2GM/V_0^2 \approx 89$ kpc.

Assuming an exponential disk of mass $4.5\times10^{10} M_{\odot}$ and radial scale length 2.5 kpc \cite{Binney2008}, the rotational velocity of the disk at the location of the sun ($r_{\odot}=8$ kpc) is approximately 160 km s$^{-1}$. Setting the total rotational velocity at $r_{\odot}$ equal to 220 km s$^{-1}$, the radial scale length of the filament from Eq. \ref{eq9} is $R\approx 8.6$ kpc. Therefore, the axis ratio of the filament is $2R/l\approx 0.19$, consistent with simulations involving warm dark matter \cite{Gao2007}. By comparison, the average axis ratio of halos in CDM simulations is around 0.6 \cite{Bailin2005}. It follows from Eq. \ref{eq6} that 92 \% of the total halo mass is contained within 50 kpc, which is consistent with data of nearby satellite galaxies and globular clusters \cite{Wilkinson1999,Sakamoto2003}. 

The velocity dispersion is related to the filament mass and velocity anisotropy parameter through Eq. \ref{eq7}. Thus, $\sigma=131$ km s$^{-1}$ and the density at the location of the sun from Eq. \ref{eq5} is $\rho(r_{\odot})\approx 0.19$ GeV cm$^{-3}$, which compares well to the currently accepted value 0.3 GeV cm$^{-3}$ \cite{Gaitskell2004}. Therefore, the total halo mass and local dark matter density are reduced considerably from the values derived using spherical halo models.

It is worth considering the possbility that neutrinos might constitute a filament of dark matter in the Milky Way galaxy. Unlike CDM particles, the existence of neutrinos has been verified experimentally. Neutrinos become free-streaming with a characteristic velocity that is approximately equal to their thermal velocity $V_{th}$, which decays as \cite{Lesgourgues2006}
$$
V_{th}\approx 150 (1+z)\left( \frac{1 \text{ eV}}{m_{\nu}} \right) \text{ km s}^{-1},
$$
where $z$ is the redshift parameter and $m_{\nu}$ is the neutrino mass. Setting the neutrino thermal velocity equal to the velocity dispersion of the filament $\sigma=131$ km s$^{-1}$, at present ($z = 0$) the neutrino mass is $m_{\nu}\approx 1.1$ eV. By comparison, gravitational lensing data of the galaxy cluster Abel 1689 supports thermal, non-relativistic, gravitating neutrinos of mass of 1.5 eV \cite{Nieuwenhuizen2009}.

\section{\label{sec:level1}Summary}
A cylindrical model for the dark matter halo of spiral galaxies has been developed. At the center the cylinder, in the plane perpendicular to the long axis, the circular velocity is constant for distances much less than the filament length and Keplerian at much greater distances. The maximum value of the circular velocity in the flat portion of the rotation curve is determined by the linear mass density of the filament. In the plane of the disk, the rotation curve is equivalent the TF profile, a model that was derived empirically to explain the observed radial velocity dispersion of the Milky Way dark halo. The cylinder model is distinguished from the isothermal sphere and NFW profile by its abrupt Keplerian decline at large distances from the galactic center. When applied to the Milky Way dark halo, the cylinder model can explain the observed radial velocity dispersion with nearly three times less mass than the NFW profile.

Assuming an axisymmetric, isothermal, steady-state filament, the rotation curve in the innermost region varies as $V_c\propto r^{1-\beta/2}$, thus establishing a relation between the functional form of the rotation curve and the velocity dispersion of the dark matter. Within observational constraints, a cylindrical model of the Milky Way dark halo appears to be consistent with hot dark matter consisting of free-streaming neutrinos with a mass of 1.1 eV.

\bibliography{bib}

\end{document}